\documentclass[prb,twocolumn,superscriptaddress, showpacs]{revtex4}

\usepackage[dvips]{graphicx}

\begin{document}

\title{$S=1/2$ chains and spin-Peierls transition in TiOCl }

\author{Alexander Seidel}
\affiliation{Center for Material Science and Engineering, Massachusetts Institute of Technology, Cambridge, MA 02139}
\affiliation{Department of Physics, Massachusetts Institute of Technology, Cambridge, MA 02139}
\author{Chris A. Marianetti}
\affiliation{Center for Material Science and Engineering, Massachusetts Institute of Technology, Cambridge, MA 02139}
\affiliation{Department of Material Science and Engineering, Massachusetts Institute of Technology, Cambridge, MA 02139}
\author{F. C. Chou}
\affiliation{Center for Material Science and Engineering, Massachusetts Institute of Technology, Cambridge, MA 02139}
\author{Gerbrand Ceder}
\affiliation{Center for Material Science and Engineering, Massachusetts Institute of Technology, Cambridge, MA 02139}
\affiliation{Department of Material Science and Engineering, Massachusetts Institute of Technology, Cambridge, MA 02139}
\author{Patrick A. Lee} 
\affiliation{Center for Material Science and Engineering, Massachusetts Institute of Technology, Cambridge, MA 02139}
\affiliation{Department of Physics, Massachusetts Institute of Technology, Cambridge, MA 02139}

\date{\today}

\begin{abstract}
We study TiOCl as an example of an $S=1/2$ layered Mott insulator. From our analysis of new susceptibility data, combined with LDA and LDA+U band structure calculations, we conclude that orbital ordering produces quasi-one-dimensional spin chains and that TiOCl is a new example of Heisenberg-chains which undergo a spin-Peierls transition. The energy scale is an order of magnitude larger than that of previously known examples. The effects of non-magnetic Sc impurities are explained using a model of broken finite chains.

\end{abstract}

\maketitle

The discovery of high-$T_c$ superconductors has generated a great deal of interest in low-dimensional spin-$1/2$ materials. Even in the absence of charge degrees of freedom, many questions remain unanswered until today. One of key importance is the question whether a two-dimensional spin system can support unbroken symmetry at zero temperature due to strong quantum fluctuations for $S=1/2$ systems. Such a spin liquid state was suggested by Anderson \cite{PWA} and termed the resonating valence bond (RVB) state . This has motivated us to search for other examples of $S=1/2$ layered materials, notably at the beginning of the transition metal series where Ti$^{3+}$ is in the $d^1$ configuration (as opposed to $d^9$ in the cuprates). From this point of view the layered compound titanium oxihalides TiOX (X=Cl, Br) appear most promising. Indeed, Beynon and Wilson \cite{WIL} reported in 1993 that the uniform magnetic susceptibility of these materials shows a number of unusual properties. According to them, the susceptibility is almost temperature independent and could not be fitted to the Curie-Weiss law, nor to any one-dimensional model. Moreover, if non-magnetic Sc ($d^0$) impurities were introduced, a large Curie tail corresponding to one spin-$1/2$ moment per Sc appears. After subtracting this Curie tail, the susceptibility was again found to be temperature independent but substantially reduced compared to the pure material. This effect was particularly striking in TiOCl and led Beynon and Wilson to propose that these materials may be examples of RVB-like states.\\
\indent We have redone the measurement of the susceptibility of TiOCl for both the pure material and in the presence of Sc, and carried out LDA and LDA+U band structure calculations. Our data are not in agreement with that of ref.\onlinecite{WIL}, especially for the pure sample. Here, we find that above a temperature of $130K$, below which the susceptibility drops abruptly, our data fits well to a nearest neighbor Heisenberg model with an exchange constant of $J=660K$.
The data for $6\%$ and $10\%$ Sc-doping is in very good agreement with a model based on broken finite linear Heisenberg-chains.\\ 
\indent In the following we describe the structure of TiOCl and discuss the possibility of 1d spin chains formed by $t_{2g}$ orbitals. The reasoning here will be justified by  band structure calculations, which feature isolated one-dimensional $d$-bands. We then present our experimental data for the pure and Sc-doped samples and fits to theoretical models. We will argue that all evidence is consistent with a picture based on quasi-one-dimensional spin-$1/2$ chains, where the pure material undergoes a spin-Peierls transition at low temperatures.\\
\begin{figure*}
\begin{center}
\includegraphics[width=7in]{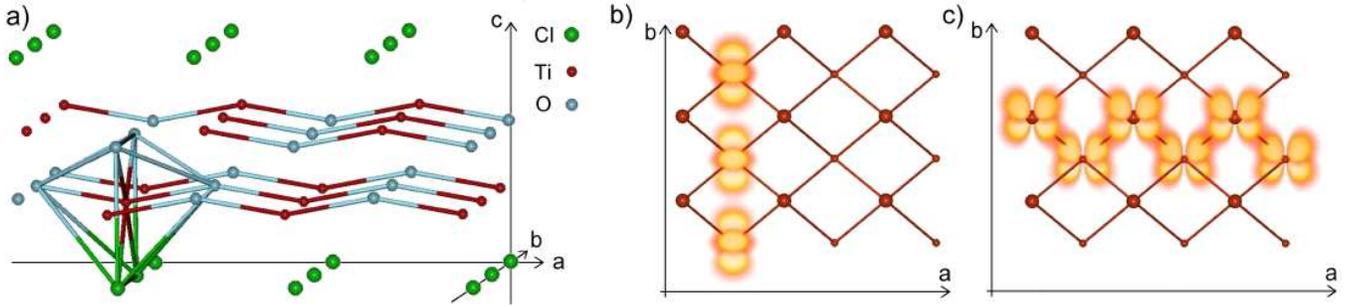}
\caption{\label{struct} a) Structure of TiOCl. A single bilayer is shown. Note that each Ti is surrounded by 4 O and 2 Cl, forming a distorted octahedron. These octahedra are corner-sharing along a and edge-sharing along b.
b+c) Top view on the lattice showing only Ti ions which form two rectangular layers. The smaller symbols are for the bottom layer. b) $d_{xy}$ orbitals forming linear chains. c) $d_{xz}$ orbitals forming zig-zag chains. Note that the plane of the orbital is tilted by $45^\circ$ out of the paper and two of its lobes point directly at the Ti ions on the adjacent layer. Another zig-zag chain is formed by the degenerate $d_{yz}$ orbital and orbital ordering will be needed to break the symmetry.}
\vspace{-1cm}
\end{center}
\end{figure*}
\indent The structure of TiOCl is of the FeOCl-type which consists of bilayers as displayed in Fig.\ref{struct}a).  The symmetry is orthorhombic with $Z=2$.
The layers repeat in the c-direction of the crystal with $c=8.03$\AA.  Cl layers mediate a weak van der Waals interaction between successive bilayers. Within each bilayer, Ti and O form two layers of buckled chains where Ti is always on the outer side with respect to the bilayer.
The O-Ti-O bond angle is $153^\circ$. Each Ti ion is surrounded by a distorted octahedron of O and Cl ions (Fig.\ref{struct}a))).
 These octahedra are formed by two O ions belonging to the same Ti-O chain, two O ions belonging to neighboring chains, and two Cl which lie on the outside of the bilayer. They are corner-sharing in the a-direction along the Ti-O chains and edge-sharing in the b-direction. Figs.\ref{struct}b) and \ref{struct}c) show a top view of the lattice showing only the Ti ions for clarity. The Ti sublattice consists of two rectangular layers with lattice parameters $a=3.79${\AA} and $b=3.38${\AA} in each layer. The top layer is shifted laterally and is displaced vertically from the bottom layer by $1.96$\AA. The shortest Ti-Ti bond length turns out to be the distance between Ti in different layers, and at $3.21${\AA} is just slightly shorter than the Ti-Ti distance along b. For orientation, the Ti ions in each layer are bridged by oxygen ions (not shown in figs.\ref{struct}b)+c)) along the a-direction, forming the buckled Ti-O chains described earlier. Note that these buckled chains (Fig.\ref{struct}a)) are introduced for the purpose of describing the crystal structure only, and should not be confused with the spin chains, which we will next propose, that are based on the electronic structure. In particular, we believe that the important exchange path is direct $t_{2g}$ orbital overlap, rather than superexchange

\begin{figure}
\samepage
\begin{center}
\includegraphics[width=8cm]{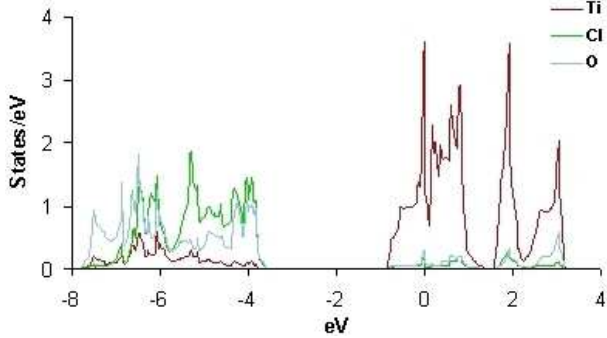}
\caption{\label{proj} LDA-DOS projected onto Ti, O and Cl orbitals}
\vspace{-1.1cm}
\end{center}
\end{figure}
 via oxygen. At each Ti site, the approximately octahedral co-ordination dictates a set of axes for the conventional  $e_g$ and $t_{2g}$ orbitals. We chose $\hat{z}=a$, and $\hat{x}$ and $\hat{y}$ axes which are rotated by $45^\circ$ relative to the $b$ and $c$ axes. With this choice of axes, the $d_{x^2-y^2}$ and $d_{z^2}$ orbitals point towards the oxygen and chlorine neighbors as required for $e_g$ orbitals. The important orbitals are the $t_{2g}$'s ($d_{xy}$, $d_{xz}$ and $d_{yz}$). The $d_{xy}$ orbitals form linear chains running along the b-axis, linking Ti ions in the same layer. This is shown in Fig.\ref{struct}b). On the other hand, Fig.\ref{struct}c) shows that the $d_{xz}$ orbitals are tilted by $45^\circ$ out of the plane and two of the lobes point directly at the Ti ions in the other layer. This is also the direction of the shortest Ti-Ti distance and we propose that these orbitals may form a zig-zag chain along the a-direction. Note that such a state is degenerate with a similar state derived from the $d_{yz}$-orbitals, where the latter are connecting a different set of pairs of neighboring chains. A state with infinite zig-zag chains is thus one of a broken two-fold symmetry between the $d_{xz}$- and $d_{yz}$-orbitals. 


\begin{figure}
\begin{center}
\includegraphics[width=8cm]{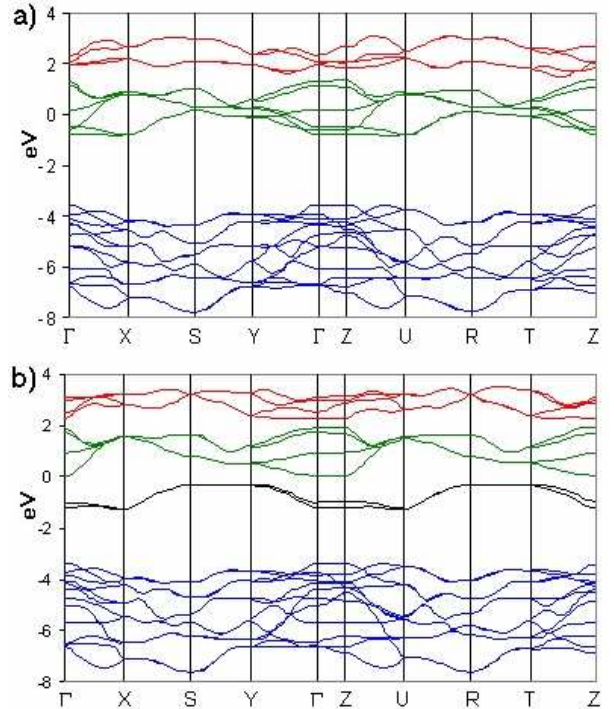}
\caption{\label{bands} a) LDA band structure calculation for TiOCl. The Brillouin zone is orthorhombic with $X=2\pi\hat{x}/a$,$Y=2\pi\hat{y}/b$,$Z=2\pi\hat{z}/c$ and S is the zone corner. The unit cell contains 2 Ti atoms. 6 bands around the Fermi level are identified as $t_{2g}$ bands well separated from the $e_g$ bands. b) LDA+U band structure calculation with split off $d_{xy}$ bands. }
\vspace{-1.0cm}
\end{center}
\end{figure}

The arguments given thus far assume that the states at the Fermi level are 
 $t_{2g}$-like electrons. In order to confirm this, and to investigate orbital 
ordering, we have performed both LDA and LDA+U \cite{ANI}
 calculations using the 
full-potential LMTO \cite{AND} method. All calculations were performed 
using the LmtART code \cite{SAV}, and the experimentally determined unit cell \cite{SCHAFER}, and ferromagnetic spin polarization. Both LDA and LDA+U calculations yielded a magnetic moment of 
$1 \mu_b$ per formula unit. In the following discussion, the band structures 
and DOS correspond to the majority spin, as the minority spin for the d-states are completely unoccupied. We begin by presenting the site-decomposed density of 
states Fig.\ref{proj}. As shown, the oxygen and chlorine p-levels form well 
separated bands from the Ti d-levels with only small hybridization between 
the two. Also, the octahedral crystal field has clearly split the d-states 
into $t_{2g}$ and $e_g$ contributions, with the Fermi level lying within the 
$t_{2g}$ peak as expected for a $d^1$ configuration. These features can also 
be seen in the band structure (Fig.\ref{bands}a)). Therefore, the assumption of $t_{2g}$-like electrons at the Fermi level is valid. However, LDA predicts this material to be metallic, and projecting the DOS onto the $t_{2g}$ orbitals indicates that all three orbitals are partially occupied (not shown). In order to go beyond the LDA and explicitly include 
the effect of strong on-site interactions which may induce orbital 
ordering, we have carried out LDA+U calculations. We performed calculations with $U=3.3eV$ and a ferromagnetic on-site exchange of $1eV$ (Fig.\ref{bands}b)). As shown, two nearly degenerate, one-dimensional bands split off from the rest of the $t_{2g}$ bands creating an insulator (note that there are 2 atoms per unit cell). These two bands are derived from the $d_{xy}$ orbitals corresponding to the linear chains in Fig.\ref{struct}b). Varying U only had a significant effect on the splitting between the occupied and unoccupied bands, not on the shape or width of the occupied bands. If the band width of 0.9 eV is identified with $4t$, where $t$ is the nearest neighbor hopping in a one-dimensional tight binding model, we can estimate the exchange constant in an effective Heisenberg model via $J=4t^2/U=714K$.

Other band structure calculations for TiOCl found in the literature \cite{KIM} that are based on a phenomenological tight binding model are not in agreement with our LDA calculations. In ref.\onlinecite{KIM}, Cl seems to play an important role at the Fermi surface and strong correlation effects leading to orbital ordering have not been included.


\begin{figure}
\begin{center}
\includegraphics[width=7.5cm]{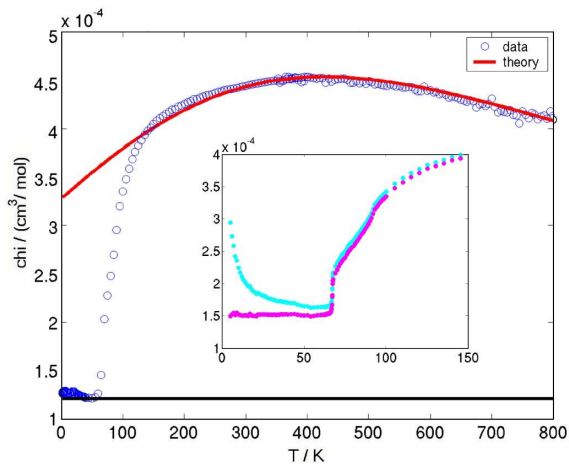}
\caption{\label{data00} Susceptibility of crushed crystals of pure TiOCl. The inset shows single crystal data with and without Curie subtraction. }
\vspace{-.9cm}
\end{center}
\end{figure}

\begin{figure}
\begin{center}
\includegraphics[width=7.3cm]{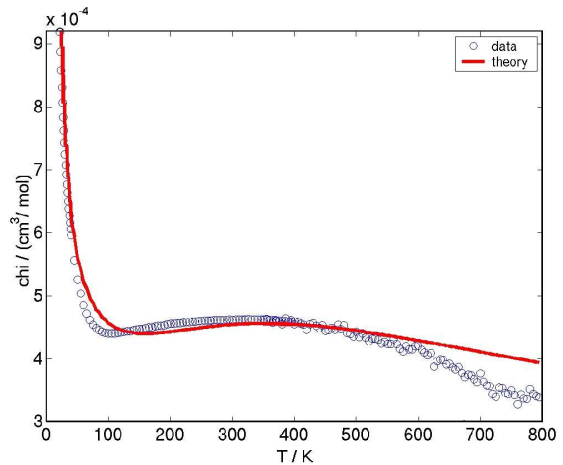}
\caption{\label{data10} Susceptibility of Sc doped TiOCl (batch A) }
\vspace{-.9cm}
\end{center}
\end{figure}

The band structure thus indeed suggests the existence of one-dimensional spin-$1/2$ chains in TiOCl, where the spins are localized in Ti $t_{2g}$-orbitals. We have repeated the measurement of the uniform susceptibility of pure {TiOCl} (Fig.\ref{data00}) and in the presence of Sc impurities (Fig.\ref{data10}), over a temperature range from a few K to 800K.

Single crystals TiOCl and Ti$_{1-x}$Sc$_x$OCl were prepared by the chemical vapor transport method using excess TiCl$_3$ as transport agent \cite{SCHAFER}. TiOCl precursor is a mixture of TiO$_2$/TiCl$_3$ (1:6 by weight) and Sc doped precursor is a mixture of Ti$_2$O$_3$/Sc$_2$O$_3$/TiCl$_3$ with 6\% and 15\% of Sc in the evacuated  sealed quartz tubes. TiOCl crystal is transported by TiCl$_3$ vapor to the cold zone with a gradient maintained at $650$C$/550$C within $20$cm.  It takes three to five days to complete the vapor transportation. The Sc doping level of the single crystal may not follow the calculated initial Sc/Ti ratio due to the uncontrolled transport mechanism with excess TiCl$_3$.  However,  we find that the thoroughly mixed and reacted powder samples have Curie constants corresponding to the amount of free Ti$^{3+}$ spins that are very close to half of the calculated initial Sc percentage.
   
The average Curie
   constants for single crystals are $5.8\pm 1.2\%$ Ti$^{3+}$ for $15\%$ Sc initial
   mixture (batch A) and $3.1\pm 0.3\%$ for $6\%$ initial Sc mixture (batch B).  It is very likely
   that the Curie constant does reflect only half of the Sc doped in the
   sample.

Our data for pure TiOCl shown in Fig.\ref{data00} were obtained from crushed crystals.They are about a factor 3 smaller than those proposed in ref.\onlinecite{WIL}. After subtraction for a small Curie tail and correction for trace amounts of ferromagnetism, we find a sharp drop in the susceptibility at low temperatures. In a picture based on 1d spin chains, such a drop must be interpreted as a spin-Peierls transition. The flat part of the curve below 50K then determines the zero-level of susceptibility of the Ti d-electrons. The single crystal data shown in the insert of Fig. \ref{data00} features a sudden drop of the susceptibility to zero at 67 K. In addition, a noticealbe inflection point appears at 95 K. The high temperature part of the data (above 130K) can be fitted to the Bonner-Fisher-curve \cite{BF}, using the nearest neighbor exchange $J$ as the only free parameter of the Heisenberg-Hamiltonian:

\vspace{-.4cm}

\begin{equation}
  H=J\sum_{i}S_i\cdot S_{i+1}
\end{equation}

\vspace{-.3cm}

where the $S_i$ are spin-$1/2$ operators.
 The solid curve in Fig.\ref{data00} shows that the data are indeed very well described by this model. Note that both the absolute magnitude and the peak position are well accounted for by a single parameter $J$, justifying the choice of zero made above.
The fit determines $J=660K$, which agrees well with the crude estimate given above. The result of Sc-doping can now be explained. In our model, we assume that Sc replaces Ti on a fraction $x$ of all Ti-sites. Sc is in a $d^0$ configuration, therefore Sc-doping can be regarded as the analogue of Zn-doping in the cuprates. The Heisenberg-chains of our model are thus broken into finite open-ended chains of an average length of $1/x$. More precisely, let $P_x(n)=nx^2(1-x)^n$ denote the probability for a site to belong to a chain of length $n$. Denoting the susceptibility per site of a finite open-ended Heisenberg chain of length $n$ by $\chi(n,T)$, we expect the measured susceptibility in the presence of Sc to be of the form:

\vspace{-.4cm}

\begin{equation}
  \chi_{Sc}(x,T)=\sum_{n=1}^\infty P_x(n)\chi(n,T)
\end{equation}

\vspace{-.2cm}

We have determined $\chi(n,T)$ by exact diagonalization for $n=1..16$ and extrapolation. As shown in Fig. \ref{data10}, this model with $x=.1$ agrees well with the experiment on batch A at all temperatures below $600K$, above which the doped sample begins to decompose. Similar agreement was achieved for batch B with x=.06. The existence of a Curie-like tail at low temperatures is easily understood in our model and is due to the presence of chains of odd length, which behave as free spin-$1/2$'s at temperatures below the gap. Since the gap decreases as $1/n$ as the chain length $n$ increases, chains of different lengths will contribute to the Curie-like tail at different temperatures. The fitting to a Curie-tail is therefore imperfect, as pointed out in ref.\onlinecite{WIL}, except at very low temperatures, where all odd chains contribute. It is clear that in the limit of small $x$ half of all chains in the system will be odd. Hence in our model, Sc doping roughly introduces free spin-$1/2$'s into the system on a 2:1 basis at very low temperatures, as conjectured before. This result is in contradiction with ref.\onlinecite{WIL} where a 1:1 correspondence is reported. Furthermore, in ref.\onlinecite{WIL} a large drop of the flat high temperature part of the susceptibility was found upon introduction of Sc into TiOCl. This again does not agree with our measurement, where the high temperature parts of the pure and doped sample have very similar values over a wide temperature regime.  We remark however that the effect of Sc doing as measured by us resembles that reported in ref.\onlinecite{WIL} for the case of TiOBr, as one would expect for two such similar chemical compounds. 
\indent Note that our model does not assume any spin-Peierls transition for the Sc-doped cases. However, we believe that the susceptibility data alone does not allow us to draw any firm conclusion as to whether the spin-Peierls transition is destroyed or not. \\
\indent We conclude that both our numerical and experimental results consistently indicate the existence of one-dimensional spin-1/2 chains in TiOCl. Above $130K$, the susceptibility of TiOCl is well described by a nearest-neighbor-exchange Heisenberg model with $J=660K$. We interpret the abrupt drop of the susceptibility at $67K$ as a transition into a spin-Peierls state. We have identified the two possible kinds of spin chains that may be present. These are linear chains along the b-axis and zig-zag chains along the a-axis of the crystal. The $t_{2g}$ orbitals that form these chains have been shown to be energetically isolated in the band structure calculation. We note that in the case of the linear chains it is clear from fig.\ref{struct}b) that the neighboring chains form a staggered pattern and the exchange coupling between neighboring chains is frustrated. Thus interchain coupling effects are likely to be weak. Furthermore, below the spin-Peierls transition, the lattice distortion associated with the dimerization is also frustrated from one chain to the next. This is not the case for the zig-zag chains. Once the degeneracy of the zig-zag direction is broken, the interchain exchange is unfrustrated and zig-zag chains can easily lock in place. Thus interchain effects are likely to be different in the two scenarios. A first order phase transition into a Spin-Peierls state as suggested by our single crystal data is expected  by Ginzburg-Landau arguments only in the frustrated scenario of the linear chains. This is also the scenario favored by the LDA+U calculation. Further experiments such as x-ray diffration are needed to confirm this picture.\\
\indent  A number of other spin-Peierls systems have been studied, notably the organic system TTF-CuBDT \cite{MORT} (where the spin resides on TTF) and CuGeO$_3$ \cite{HASE}. The exchange constants and transition temperatures are $J=77K$, $T_c=12K$ for TTF-CuBDT and $J=88K$, $T_c=14K$ for CuGeO$_3$ and the interchain coupling is unfrustrated in both cases. The energy scale for TiOCl is close to an order of magnitude higher. The relatively large J and the weak Jahn-Teller effect for $t_{2g}$ orbitals both tend to minimize carrier localization effects making TiOCl an attractive candidate for doping. For large doping $x$ where $xt$ exceeds the spin-Peierls energy scale, the doped holes will destroy the spin-Peierls state, resulting in a Luttinger liquid. For smaller hole concentration the spin-Peierls local order survives and the spins remain in singlet pairs.  Coherent motion of doped carriers may lead to unconventional superconductivity, as in the original RVB scenario. These possibilities are certainly worthy of further investigations.\\

\vspace{-1.2cm}

\begin{acknowledgments}

\vspace{-.5cm}

We thank Gil Lonzarich for bringing ref.\onlinecite{WIL} to our attention and T. Senthil for insightful discussions. The work of A.S., F.C.C. and P.A.L. was supported by the MRSEC program of the NSF under award number DMR 98-08941. The work of C.A.M. and G.C. was supported by the DOE contract number DE-FG02-96ER45571.  
\end{acknowledgments}

\end{document}